\documentclass{svproc}
\usepackage{url}
\usepackage{amsmath}
\usepackage{graphicx,psfrag,epsf}
\usepackage{enumerate}
\usepackage{amssymb}
\usepackage{bbm}
\usepackage{amsfonts}
\usepackage{subfig,xcolor}
\usepackage{wrapfig}
\usepackage{mathtools}

\begin{document}
\mainmatter              % start of a contribution
\title{Uncertainty Quantification for Agent Based Models: A Tutorial}
\titlerunning{UQ for ABMs}  % abbreviated title (for running head)
\author{Louise Kimpton \and Peter Challenor \and
James Salter}
%
%\authorrunning{Louise Kimpton et al.} % abbreviated author list (for running head)
%\tocauthor{Louise Kimpton, Peter Challenor, James Salter}
%
\institute{University of Exeter, UK\\
\email{l.m.kimpton@exeter.ac.uk}}

\maketitle              % typeset the title of the contribution

\begin{abstract}

We explore the application of uncertainty quantification methods to agent-based models (ABMs) using a simple sheep and wolf predator-prey model. This work serves as a tutorial on how techniques like emulation can be powerful tools in this context. We also highlight the importance of advanced statistical methods in effectively utilising computationally expensive ABMs. Specifically, we implement stochastic Gaussian processes, Gaussian process classification, sequential design, and history matching to address uncertainties in model input parameters and outputs. Our results show that these methods significantly enhance the robustness, accuracy, and predictive power of ABMs.
% We would like to encourage you to list your keywords within
% the abstract section using the \keywords{...} command.
\end{abstract}
\section{Introduction}
Agent-based models (ABMs) are powerful computational tools designed to simulate the actions and interactions of autonomous agents within a defined environment. These agents, which represent individuals such as people, animals, or organizations, operate according to a set of rules. The agents can adapt to changes in their environment or the behaviour of other agents, which allows researchers to study the emergent behaviour of complex systems as well as observe both individuals and interactions.  ABMs are particularly valuable in fields such as epidemiology, economics, ecology, and social sciences, where individual heterogeneity and local interactions play a crucial role in shaping system dynamics.

However, the complexity and stochastic nature of ABMs introduces significant challenges in terms of uncertainty quantification (UQ) \cite{Craig2001,Currin1991,Haylock1996}.  Uncertainty in ABMs can arise from various sources, including the randomness inherent in agent behaviours, variability in model parameters, and the initial conditions of the system.  High-dimensional and non-linear interactions among agents can amplify these uncertainties, making it difficult to predict the overall system behaviour with confidence.  Additionally,  increased agents results in increased computation time of the model, making it difficult to run at many initial input settings. With the increased use of high powered computing comes the ability to build larger ABMs with more complex rule sets and higher numbers of agents \cite{Ozik2021}.  Although large scale computers can help reduce run time through parallelisation, large ABMs can still be very computationally expensive: UQ techniques, including emulation, then become vital in prediction whilst also addressing uncertainty.

Addressing these uncertainties is crucial for enhancing the reliability and predictive power of ABMs. Effective UQ techniques allow researchers to assess the confidence in their model predictions, identify the most influential parameters,  guide the design of robust decision-making strategies, and calibrate models \cite{Fadikar2018,Andrianakis2017,McCulloch2022}.  By systematically quantifying and managing uncertainties, researchers can improve model validation, ensure the robustness of simulation outcomes, and provide more reliable guidance for policy and decision-making \cite{Lempert2019}.

In this paper, we explore various methodologies for uncertainty quantification in agent-based models, emphasising their importance in validating model outputs and improving model robustness.  These methods include stochastic emulation, classification, sequential design, and calibration in the form of history matching. As a tutorial example we will be showcasing UQ techniques using the simple wolf and sheep predator-prey ecosystem model.  Exemplar code can be found at https://github.com/LMKimpton2/UQforABMs.

\section{Emulation}
When dealing with computationally expensive computer models, emulators \cite{Currin1991,Sacks1989,Santner2003} are typically used as statistical surrogates that approximate the relationship between the inputs and outputs of the complex model.  An emulator is a computationally fast stochastic approximation to, in this case, the ABM.  The expectation of the emulator produces a prediction, and the variance gives a measure of the extra uncertainty coming from the emulation.  Emulators often only require a relatively small set of model runs, known as the training data, and by learning from these runs, the emulator can predict the outputs for new, unseen inputs with minimal computational effort. This approach significantly reduces the computational burden associated with running detailed simulations, allowing for more extensive and efficient exploration of the model.

Assume the computer model, or ABM, is represented by the mathematical function $f : \mathcal{X} \mapsto \mathbb{R}$ with $\textbf{X} := (\textbf{x}_{1}, \ldots, \textbf{x}_{n})^{T} \in \mathcal{X}$ being a set of $n$ inputs to the model in $p$ dimensions. The function $f(\cdot)$ hence maps the inputs $\textbf{X}$ to their associated outputs $\textbf{y} = (f(\textbf{x}_{1}), \ldots, f(\textbf{x}_{n}))^{T}$.  A common choice of emulator is a Gaussian process (GP) \cite{Rasmussen2006,Santner2003}.  In a GP, any finite collection of function values follows a multivariate Normal distribution. A GP, $Z$ is typically considered as a sum of two processes:
\begin{equation}
Z = m(\textbf{x}) + \varepsilon(\textbf{x}), 
\end{equation}
where $m(\textbf{x})$ represents a deterministic global response surface behaviour, and $\varepsilon(\textbf{x})$ is a correlated residual process capturing local input dependent deviation from the global response surface.  We can hence model $f$ as a Gaussian process $Z$ with training data $\{\textbf{X}, \textbf{y}\}$ such that:
\begin{equation} \label{EQ3}
Z(\textbf{x}) \sim GP(m(\textbf{x}), k(\textbf{x}, \textbf{x}') + \tau^{2}).
\end{equation}
The GP is specified by a mean function $m(\cdot)$, which represents the expected value of the function at any point, and a covariance function (or kernel) $k(\cdot, \cdot)$, which describes the correlation between function values at different points. The choice of kernel encodes assumptions about the smoothness and variability of $f$.  Here, $\tau^{2}$ is known as the nugget parameter, either used for numerical stability or for allowing stochastic variation in the outputs.  Using standard multivariate Normal distribution results, we can predict the function $f$ at a new location $\textbf{x}$ with posterior mean $m^{*}(\cdot)$ and variance $k^{*}(\cdot)$ given as: 
\begin{equation}
\begin{split}
m^{*}(\textbf{x}) &= m(\textbf{x}) + k(\textbf{x},\textbf{X}) (k(\textbf{X}, \textbf{X}) + \tau^{2})^{-1} (\textbf{y} - m(\textbf{x})), \\
k^{*}(\textbf{x}) &= k(\textbf{x}, \textbf{x}) + \tau^{2} - k(\textbf{x},\textbf{X}) (k(\textbf{X}, \textbf{X}) + \tau^{2})^{-1} k(\textbf{X},\textbf{x}).
\end{split}
\end{equation}

Agent based models are typically stochastic models since they provide different outputs when run at the exact same input parameters. This makes the standard noiseless Gaussian process emulator used in deterministic problems unsuitable; we can no longer assume that the intrinsic variance is constant across the input space. The nugget term in Equation \ref{EQ3} represents this intrinsic variance of the simulator and we hence define $\tau^{2}(\textbf{x})$ to be dependent on the input parameters $\textbf{x}$. This is known as heteroskedasticity.

For all stochastic models, running several replicates of the model at the same input values becomes important since we gain valuable information from the replicates about the true variance of the process.  If there are enough replicates at each design point, then we can observe an approximation to both $f$ and $\tau^{2}$ by calculating both the sample means and variances.  We can then model both the mean response and the intrinsic variance with independent Gaussian process, using the sample quantities as training data \cite{Goldberg1998,Kersting2007,Binois2018}:
\begin{equation} \label{Eq1}
\begin{split}
Z(\textbf{x}) &\sim GP(m(\textbf{x}), k(\textbf{x}, \textbf{x}') + \tau^{2}(\textbf{x})) \\
\text{log}(\tau^{2}(\textbf{x})) &\sim GP(m_{\tau}(\textbf{x}), k_{\tau}(\textbf{x}, \textbf{x}')).
\end{split}
\end{equation}
The logarithm of the intrinsic variance is modelled as a Gaussian process instead of the intrinsic variance itself to enforce positivity.  

If there are not enough replicates at each input, then the sample variance does not become a good estimate for the intrinsic variance. A different strategy is to treat the values of the intrinsic variance at the training input points as a set of additional parameters which need to be learnt. Goldberg et al. (1998) \cite{Goldberg1998} does this in a fully Bayesian way, using Markov chain Monte Carlo. Point estimates can instead be used for these variance parameters, which loses the complete quantification of uncertainty, but provides increased computational efficiency.  Binois et al. (2018) \cite{Binois2018} have produced a more computationally efficient method known as hetGP, which is now a widely known method to fit heteroskedastic Gaussian processes (GPs with non-constant variances). They solve the computational issues by using Woodbury identities to reduce the number of latent variables.  The estimates for $\tau^{2}(\textbf{x})$ are assumed without error and so no uncertainty from the intrinsic estimates is included in predictions. Although this means that the uncertainties are underestimated, the effect is not major. It is also important to note here that replicates are not required to fit a HetGP. The authors give details on how it can be more useful to run the model at inputs which are very close together. In this case, we learn both about the mean response and the intrinsic variance simultaneously.

\subsection{Wolf and Sheep Predator Prey ABM}
We apply UQ methodology to the simple sheep and wolf predator-prey ecosystem model in NetLogo.  In the model, the sheep wander randomly and the wolves look for sheep to eat.  We assume grass is infinite in the model and hence eating or moving does not change the sheep's energy levels.  Each step costs the wolves energy where they must eat sheep to replace energy, otherwise they die out.  The system is unstable if it results in extinction for either the sheep or wolves. 

To keep the example simple and easy to visualise, we restrict the model to have two input variables: sheep reproduction rate $x_{1} \in [0,1]$ and wolf reproduction rate $x_{2} \in [0,1]$. We fix all other parameters at default values (initial number of sheep: 100, initial number of wolves: 50, wolves gain from food: 20). The ABM records the number of both sheep and wolves at each time step.  There are methods to jointly model outputs in a trajectory \cite{Fadikar2023}, as well as methods to model the time series as a high dimensional output,  performing dimensionality reduction techniques such as principal component analysis before emulation \cite{Higdon2008}.  This is outside the scope of this article,  and we instead define the output $y$ of the model to be the time to wolf extinction. 

It is important that the training set $\{\textbf{X}, \textbf{y}\}$,  of the Gaussian processes is space filling across the design space to ensure all of the space is represented.  Latin hypercube designs (LHD) \cite{Ohagan2006} are commonly used to sample a multi-dimensional space efficiently. In LHD, the range of each variable is divided into equal intervals, and samples are taken such that each interval of each variable is sampled exactly once. This ensures that the samples are spread out more evenly across the entire input space, providing a more representative and comprehensive exploration compared to simple random sampling.

\begin{figure}[ht]
\centering
\includegraphics[scale=0.5]{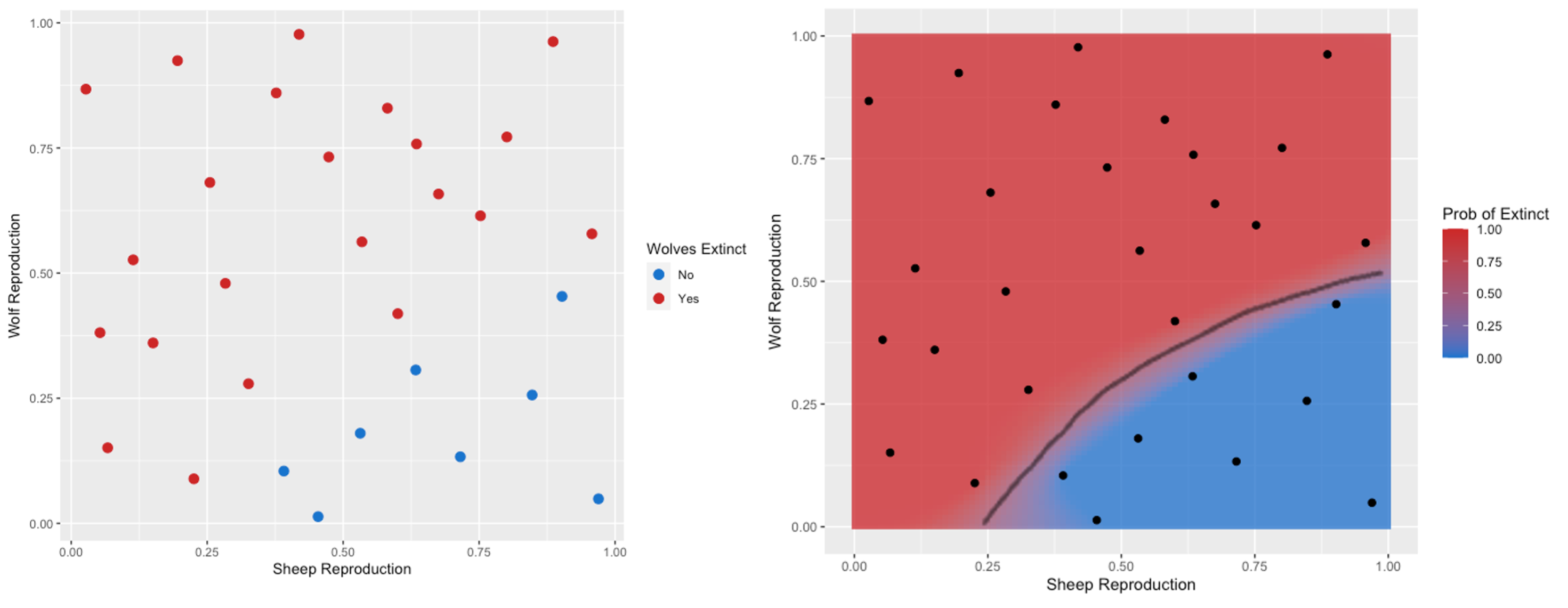}
\caption{Left: Initial maximin Latin hypercube used to obtain training data to fit the Gaussian process. Red dots are locations where a model output (wolf extinction) exists and blue points are where a model output does not exist. Right: classification prediction using GP classification. Black line gives the hard boundary and fill colour gives the probability that a model output exists.}
\label{Pic1}
\end{figure}

We have two inputs to the ABM, $\textbf{X} = (x_{1}, x_{2}) \in [0,1]^{2}$, and choose an initial design using a maximin Latin hypercube with 30 points. The maximin criteria chooses a design that maximises the minimum distance between points.  The Latin hypercube design is presented in Figure \ref{Pic1}, which we then run through the ABM to obtain the corresponding output vector $\textbf{y}$. Since the model is stochastic,  we run the Latin hypercube at 10 replicates. 

As shown by the coloured dots in Figure \ref{Pic1}, the ABM does not always produce a model output. At certain locations in the input space (blue points) the wolves fail to reach an extinction point before the model terminates. This typically occurs when there is a large reproduction rate for sheep, but a very low reproduction rate for wolves. We then find that the sheep reproduce exponentially, and the wolves have sufficient energy supply. Since this area of input space has no model output, we must first consider classification methods to fully define where the model is valid.

\subsection{Classification}
Based on the initial runs of the model in Figure \ref{Pic1}, we now aim to map the regions where wolves will and will not go extinct.  Classification methods can be used to split the input space into the separate regions according to their associated outputs.  A common method for classification is logistic regression \cite{Hilbe2009,Kleinbaum1994},  which models the probability that a given input belongs to a particular class.  For each point in the input space, it outputs a value between 0 and 1, which is then thresholded to classify the point into one of two categories. Similarly,  Gaussian process classification \cite{Rasmussen2006,Nickisch2008} models the probability of belonging to a class, but instead uses a GP to model a latent function, which is then squashed through a sigmoid function to produce the probabilistic outputs. Since the exact posterior distribution is not analytically tractable, an approximation must be used to perform inference. This is due to the non-Gaussian nature of the likelihood function (a Bernoulli likelihood for binary classification). Unlike regression, where the likelihood is Gaussian and leads to a Gaussian posterior, classification involves a non-linear transformation, making the posterior distribution non-Gaussian. To manage this, approximation methods such as Laplace approximation, expectation propagation or variational inference \cite{Rasmussen2006} are used to estimate the posterior distribution. After estimating the posterior distribution, we can then get predictions of region membership for other input points by thresholding the probabilistic outputs at 0.5, or by taking draws from a Bernoulli distribution. Improvements have been made by \cite{Kimpton2020} who are able to to find a clean boundary between regions.

By taking a GP classification approach, we are able to estimate the classification for any point in the input space. This means that we are able to predict, with uncertainty, whether the wolves are likely to go extinct without running the ABM itself. This is particularly useful if we would like to run the model at further input points so that we are confident that the model will produce an output. If we just have the goal of fitting a GP to model the time to wolf extinction,  then it is unlikely that we would want to waste computational or financial resources running our ABM in the region where wolves do not go extinct, as this will give us no extra information.  Alternatively, if we were more interested in locating the boundary between whether wolves go extinct or not, then we could run more points near the boundary to improve the classification prediction.

The right plot in Figure \ref{Pic1} shows the output of GP classification using the method described in \cite{Kimpton2020}. The black line gives the predicted hard boundary between regions, whilst the colour scale shows the probability of the wolves going extinct for each input location. 

\subsection{Stochastic Gaussian Processes}

\begin{figure}[ht]
\centering
\includegraphics[scale=0.4]{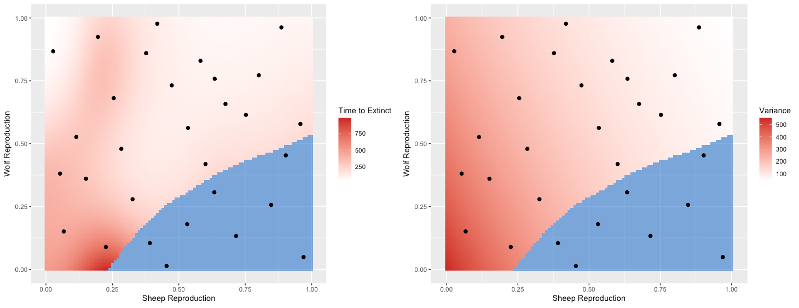}
\caption{Output of fitting a stochastic Gaussian process to the wolf and sheep predator-prey model.  The posterior mean predicted across the input space is given on the left, and the posterior variance is given on the right.}
\label{Pic2}
\end{figure}

After finding the classification estimate, we then fit an emulator to the initial data points where a quantitative model output exists.  Since the sheep and wolf predator-prey model is stochastic, deterministic Gaussian processes are no longer suitable and we must use a heteroskedastic Gaussian process as described in Section 2. Since we have few replicates at each of our design points, we have chosen to fit a stochastic Gaussian process using the HetGP package \cite{Binois2018}. The output of this is given in Figure \ref{Pic2} where the predicted mean response is given in the left hand panel, and the predicted variance is given in the right hand panel.  The expected time to extinction is highest in the bottom left hand corner where both the sheep and wolf reproduction rates are low. The rest of the output space is then fairly constant at around time 100-200 apart from a ridge when the sheep reproduction rate is around 0.25. The time to wolf extinction is predicted to be higher here regardless of the wolf reproduction rate.

The right hand panel shows the expected variance of wolf time to extinction, showing that the intrinsic variability of the model does indeed depend on the inputs. The variance is much larger in the bottom left hand corner where both the reproduction rates are low. 

\section{Sequential Design}

The accuracy of the fit of any emulator is greatly dependent on the design of the training input points \cite{Sacks1989,Santner2003,Simpson2001}. It is important that the location of these points in the input space are space filling, but also capture the important features of the physical system. This is more commonly known as the trade-off between exploration and exploitation \cite{Garud2017}.  The Latin hypercube used in Section 2 is an example of a one-shot method where all design points are chosen at once and exhaust the computational budget immediately.  A sequential design method instead selects samples one (or more) at a time using information from the emulator or existing points.  Sequential designs allow us to stop the computationally expensive sampling process as soon as the emulator reaches an acceptable level of accuracy. There is also an increased chance of exploitation, where more samples are selected in interesting areas of input space (e.g. minimums and maximums).

Many examples of sequential designs aim to minimise the predictive variance of the Gaussian process, or the mean squared error. The predictive variance is seen as an estimation of the real prediction error which increases away from existing data points. We then place new points in these areas to promote a space-filling design. Examples of current widely used design methods include minimum predictive uncertainty or mean squared error \cite{Sacks1989}, maximum entropy \cite{Shewry1987}, mutual information \cite{Krause2008}, expected squared leave-one-out cross-validation, and expected improvement \cite{Mohammadi2022}.  

For stochastic models, we are interested in learning both the mean and variance across the input space.  At every iteration we have the choice to either select a new point to improve the mean estimate, or to select a point to replicate at from the existing input design. Replicates will have a greater influence on the intrinsic variance estimate. A well known method for sequential design for stochastic models is based on reducing the integrated mean squared prediction error (IMSPE) developed by Binois et al.  (2019) \cite{Binois2019} with HetGP.  The approach starts with a space-filling design with an initial allocation of replicates at each point. A GP is fitted to this design using HetGP and  the IMSPE of the design is estimated. At each iteration, either a new point (from a set of candidates) or a replicate is included in the new design if it minimises the new potential IMSPE.

\begin{figure}[ht]
\centering
\includegraphics[scale=0.4]{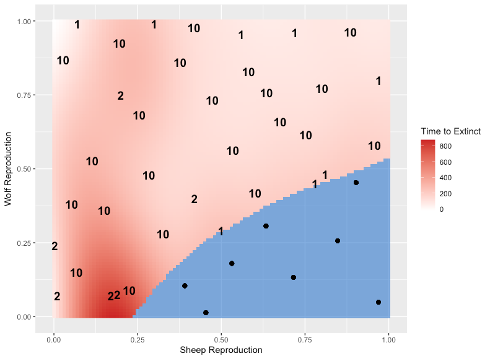}
\caption{Illustration of the sequential design method by \cite{Binois2019} for the wolf and sheep predator-prey agent based model. Plot shows the predicted mean output after 20 iterations of the design algorithm. The number at each point shows the total number of replications performed at that location after performing the sequential design.}
\label{Pic3}
\end{figure}

The plot in Figure \ref{Pic3} shows the output of applying the IMSPE sequential design method \cite{Binois2019}. We have sampled 20 extra points where the number represents the total number of replicates performed at that location (initial design points were run at 10 replications). Points in the lower left hand corner have been included based on exploitation, whilst the remaining points have been included to ensure the design is space filling.  More replicates have also been placed in the lower left hand corner to help learn the increased stochastic variance in this location. We have updated the GP using the new design and predicted across the input space.  Comparing the updated posterior mean to that with the original training data in Figure \ref{Pic2}, we observe an improvement in the prediction accuracy. The normalised root mean square error (NRMSE) calculated against a test data set shows a reduction by 0.162.

\section{History Matching}

History matching \cite{Craig1996,Vernon2010} is a form of calibration that attempts to find input parameter values to achieve consistency between observations and the agent based model. Unlike traditional calibration that provides a full probabilistic description of the input parameters, the goal of history matching is to identify the regions of input space that correspond to acceptable matches. This is performed by ruling out the implausible regions iteratively in waves. 

The GP is specified to represent the relationship between an observation $z$ and the model $f$ as:
\begin{equation}
z = f(\textbf{x}) + d + e,
\end{equation}
where $d$ represents the discrepancy between the model and the true physical system and $e$ is the observation error. Non-consistent regions of space are ruled out using an implausibility metric. The implausibility function $I(\textbf{x})$ is defined as a measure of the distance between the output of the model at $\textbf{x}$ and an observation $z$:
\begin{equation}
I(\textbf{x}) = \frac{|z - \mathbb{E}(f(\textbf{x}))|}{\sqrt{\text{Var}(e) + \text{Var}(d) + \text{Var}(f(\textbf{x}))}}
\end{equation}
Here,  $\mathbb{E}(f(\textbf{x}))=m^{*}(\textbf{x})$ and $\text{Var}(f(\textbf{x})) = k^{*}(\textbf{x}, \textbf{x})$ are the posterior mean and variance of the Gaussian process, $\text{Var}(e)$ is the variance of the observation error, and $\text{Var}(d)$ is the variance of the model discrepancy.

For model output $f(\textbf{x})$ and observation $z$, large values of the implausibility imply that, relative to our uncertainty, the predicted output of the computer model at $\textbf{x}$ is far from the observation. A threshold is chosen so that any implausibility value greater than the threshold is deemed implausible, and the corresponding $\textbf{x}$ is discarded. The remaining parameter space is termed as Not Ruled Out Yet (NROY).

\begin{figure}[ht]
\centering
\includegraphics[scale=0.47]{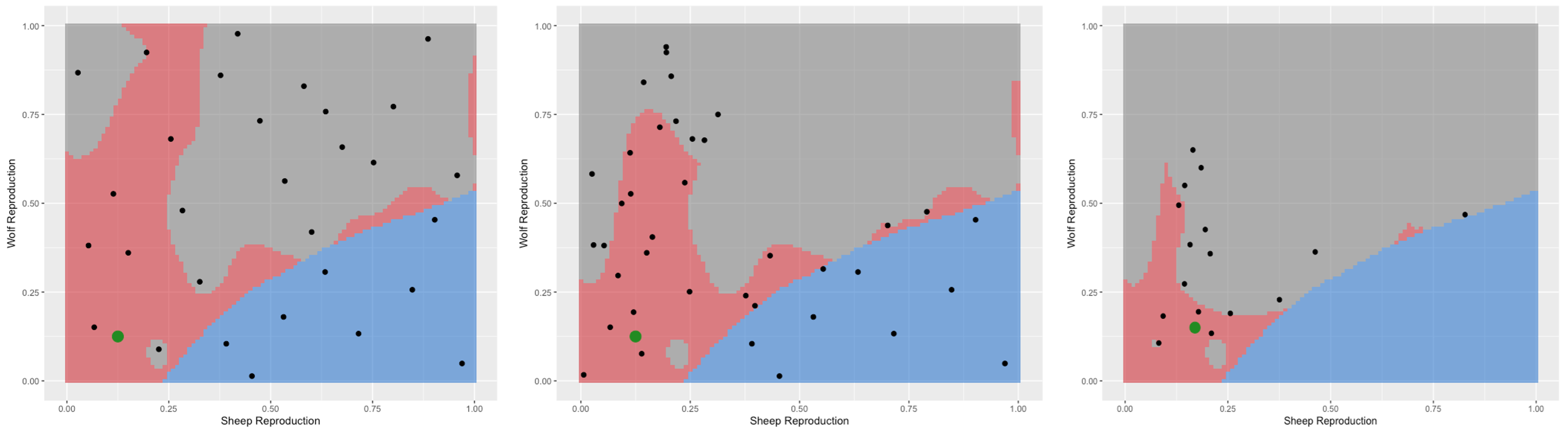}
\caption{Output after three waves of history matching (left to right).  The green point is the observation, the red region is the not ruled out yet (NROY) space and the grey region is the ruled out space deemed implausible when compared with the observation. After each wave, the black points are the new input points selected from the NROY space.}
\label{Pic4}
\end{figure}

History matching is typically performed in waves to iteratively shrink the input space. With each iteration, we set up an initial design, we run this design through the ABM, and build a GP with the training data. We then use this to rule out space, and then generate a new design from the input space that has not yet been ruled out.  History matching is applied to the sheep and wolf ABM, and the reduced space is given in Figure \ref{Pic4} with respect to an observation shown in green. Three waves of history matching have been performed, and after each wave the input space that is deemed implausible with the observation gets larger with each wave. 

\section{Conclusion}

In conclusion, the use of uncertainty quantification methods such as stochastic Gaussian processes, Gaussian process classification, sequential design, and history matching significantly enhances the robustness and predictive power of agent-based models (ABMs). By applying these methods to a simple sheep and wolf predator-prey ABM, we have demonstrated their utility in systematically addressing uncertainties inherent in model parameters and outputs. These techniques not only improve the accuracy and reliability of ABM predictions but also provide a structured framework for model validation and refinement. Future work can build upon these methods to tackle more complex systems, ensuring that ABMs remain a vital tool in understanding and simulating dynamic interactions.


\end{document}